\documentclass[11pt]{article}

\usepackage{aas_macros,amsmath,amssymb,comment,cite,esint,graphicx,mathtools}
\usepackage[margin=.8in,letterpaper]{geometry}
\usepackage[colorlinks=true]{hyperref}
\usepackage[affil-it]{authblk}
\usepackage{subcaption}
\usepackage[utf8]{inputenc}
\usepackage{mathrsfs}
\usepackage{appendix}
\usepackage{amssymb}
\usepackage{float}
\usepackage{color}
\usepackage{cite}
\usepackage{hyperref}
\hypersetup{pageanchor=false}
\usepackage{indentfirst}
\usepackage{url}
\usepackage{float}
\usepackage{caption}
\usepackage[numbers,square,comma,sort&compress,merge]{natbib}
\usepackage{esint}
\usepackage{overpic}
\usepackage{graphicx}
\usepackage{epsf,amsmath,bbold,amsfonts,stmaryrd}
\usepackage{textcomp}
\usepackage{ulem}
\usepackage{tikz}

\numberwithin{equation}{section}
\let\originalleft\left
\let\originalright\right
\renewcommand{\left}{\mathopen{}\mathclose\bgroup\originalleft}
\renewcommand{\right}{\aftergroup\egroup\originalright}
\mathcode`\*="8000
{\catcode`\*=\active\gdef*{\mathclose{}\,\mathopen{}}}

\newcommand{\td}[1]{\tilde{#1}}

\newcommand{\be}{\begin{equation}}
\newcommand{\ee}{\end{equation}}
\newcommand{\bea}{\setlength\arraycolsep{2pt} \begin{eqnarray}}
\newcommand{\eea}{\end{eqnarray}}
\newcommand{\nn}{\nonumber}

\newcommand{\mO}{{\mathcal O}}
\newcommand{\mI}{{\mathcal I}}
\newcommand{\mP}{{\mathcal P}}
\newcommand{\mQ}{{\mathcal Q}}
\newcommand{\mU}{{\mathcal U}}

\newcommand{\md}{\mathrm{d}}

\def\a{\alpha}
\def\b{\beta}

\def\d{\delta}
\def\D{\Delta}
\def\f{\frac}
\def\g{\gamma}

\def\lm{\lambda}

\def\m{\mu} 
\def\n{\nu} 
\def\nn{\nonumber}
\def\pl{\partial}
\def\p{\phi} 
\def\td{\tilde}

\def\t{\theta}

\def\k{\kappa}

\def\be{\begin{equation}}
\def\ee{\end{equation}}
\def\bag{\begin{aligned}}
\def\eag{\end{aligned}}
\def\bea{\begin{eqnarray}}
\def\eea{\end{eqnarray}}
\def\ba{\begin{array}}
\def\ea{\end{array}}

\def\bc{\begin{center}}
\def\ec{\end{center}}

\begin{document}
\title{Polarization Patterns of the Hotspots Plunging into a Kerr Black Hole}
	
\author{Bin Chen$^{1,2,3}$\thanks{E-mail: chenbin1@nbu.edu.cn}~, Yehui Hou$^{2}$\thanks{E-mail: yehuihou@pku.edu.cn}~, Yu Song$^{2}$\thanks{E-mail: songyuphy@gmail.com}~, Zhenyu Zhang$^{2}$\thanks{E-mail: jerryzhang98@hotmail.com}~}
\date{}
	
\maketitle
\vspace{-10mm}

\begin{center}
{\it

$^1$ Institute of Fundamental Physics and Quantum Technology,\\ \& School of Physical Science and Technology, Ningbo University, Ningbo, Zhejiang 315211, China\\\vspace{4mm}

$^2$ School of Physics, Peking University, No.5 Yiheyuan Rd, Beijing
100871, P.R. China\\\vspace{4mm}

$^3$ Center for High Energy Physics, Peking University,
No.5 Yiheyuan Rd, Beijing 100871, P. R. China\\\vspace{4mm}

}
\end{center}

\vspace{8mm}

\begin{abstract}

The multi-wavelength polarized light signals from supermassive black holes have sparked many studies on polarized images of accretion disks and hotspots. However, the polarization patterns within the innermost stable circular orbit (ISCO) region remain to be explored. In this study, we focus on two specific types of orbits, namely the plunging geodesics inward from the ISCO and homoclinic geodesics, to uncover the polarization features associated with non-circular motion in a Kerr spacetime. For an on-axis observer, we specifically develop an approximate function to describe gravitational lensing along the azimuthal direction and establish a simplified synchrotron emission model. Based on these, we analyze the time-integrated polarized images of hotspots and their Stokes parameters. Moreover, we explore the polarized image of the plunging region within a thin accretion disk. 

\end{abstract}

\maketitle

\newpage
\baselineskip 18pt

\section{Introduction}

Electromagnetic observations of supermassive black holes have yielded significant results in recent years. Whether it be the millimeter-band images of Sagittarius A* (Sgr A*) and M87* captured by the Event Horizon Telescope (EHT) \cite{EventHorizonTelescope:2019dse, EventHorizonTelescope:2022wkp} or the astrometry of near-infrared (NIR) flares from Sgr A* reported by the GRAVITY collaboration \cite{GRAVITY:2018det}, the observations prompt a deeper exploration of black hole physics. The findings from EHT and GRAVITY encompass the information of polarization, offering more insights into the surrounding astronomical environments near the black holes \cite{EventHorizonTelescope:2021bee,EventHorizonTelescope:2024vii,GRAVITY:2023avo}. The observed polarization angles, stemming from synchrotron radiations, encode the details about the plasma properties and magnetic field structures at the horizon-scale \cite{EventHorizonTelescope:2021srq,EventHorizonTelescope:2024viii,Ricarte:2022sxg}, aiding in the investigation of strong field physics \cite{Karas:2021ltz,Ricarte:2022wpd,Chael:2023pwp,Vincent:2023sbw,Zhang:2023cuw}.

The study of NIR flares near Sgr A* is of particular importance in understanding the dynamics of the Galactic Center. A simple yet effective approach to analyze these flares is to model them as time-varying images of a hotspot. This hotspot is a compact, high-temperature, neutral plasmoid, whose internal electrons emit synchrotron radiations \cite{Broderick:2005my, Meyer:2006fd, Trippe:2006jy, GRAVITY:2020lpa, Vos:2022yij, Huang:2024wpj, Antonopoulou:2024qco}. One plausible mechanism for generating such a plasmoid is via magnetic reconnection, a process in which magnetic field lines are severely curved and undergo reconnections to dissipate their energy \cite{SP1, Petschek, rela-SP, yamada2009, Aimar:2023kzj}. The released energy drives the plasmoids and expels them from the reconnection region. The radiation characteristics of these plasmoids render them the compelling candidates for the flares  \cite{Aimar:2023kzj}. The observations of polarized light curves at both NIR and millimeter wavelengths support the hotspot model for Sgr A* \cite{Meyer:2006fd,Vos:2022yij,Wielgus:2022heh,Yfantis:2023wsp,GRAVITY:2023avo}.

In analytical research, a ring model has been proposed to study the polarized images of hotspots around a Schwarzschild black hole, as outlined in \cite{EventHorizonTelescope:2021btj}. This model assumes an emitting circular ring, interpreted as the accumulation of an orbiting hotspot over time \cite{Rosa:2023qcv}. By modulating the locally defined magnetic field, one can generate a variety of polarizations in this model. By superimposing rings of different radii, the model can also produce a simplified version of the polarized image of an accretion disk. The ring model has since been extended to more general spherically symmetric spacetimes \cite{Claros:2024atw} and the Kerr spacetime \cite{Gelles:2021kti}. Using this analytic model, it has been demonstrated in \cite{Emami:2022kci} that the twisty morphology observed in linear polarization is primarily dictated by the magnetic field structure, with the Doppler boosting and the gravitational lensing playing subdominant roles. Additionally, a more realistic analytical model of accretion near a Schwarzschild black hole was developed in \cite{Loktev:2021nhk}. Further investigations have explored the influence of different spacetime geometries on the polarization patterns predicted by the ring model \cite{Liu:2022ruc, Qin:2022kaf, Delijski:2022jjj, Guo:2024bzq}.

There exist several unexplored avenues in the current research on the polarized images. One of the most challenging tasks is to explore the horizon-scale magnetic field, which is intricately linked to the background magnetofluid \cite{EventHorizonTelescope:2021srq}. Traditionally, the strength and the orientation of the magnetic field have been treated as phenomenological parameters in studies. Moreover, while the ring model captures the basic characteristics of orbiting hotspots or disk emission, it is not well-suited for the region within the Innermost Stable Circular Orbit (ISCO). The inner-ISCO region of an accretion flow, referred to as the plunging region, characterized by the streamlines that can be roughly approximated by a cluster of plunging geodesics, could make significant contributions to multi-wavelength observations. These observations could encompass the inner shadow observations by the EHT \cite{Chael:2021rjo} and the X-ray spectra of stellar mass black holes \cite{Machida:2002ub, Zhu:2012vf, Wilkins:2020pgu}. Furthermore, for non-circular hotspots, particularly those perturbed from unstable orbits, their astrometry and polarization patterns may diverge largely from those of circular hotspots. This divergence presents new avenues for theoretical exploration into the flares observed from Sgr A* and potentially unlock novel insights into their nature.

Previous studies have shown that the trajectories of timelike geodesics perturbed from unstable circular orbits or the ISCO can be expressed in terms of elementary functions \cite{Levin:2008yp}, enabling the purely analytical generation of hotspot images. In this work, we extend the study of imaging and polarization of circular orbits \cite{EventHorizonTelescope:2021btj, Gelles:2021kti} to non-circular orbits in a Kerr spacetime. This extension utilizes the trajectory integrals outlined in \cite{Levin:2008yp, Mummery:2023tgh}. For simplicity, we place the distant observer along the black hole's rotation axis in most cases, allowing us to derive analytical expressions for gravitational lensing. Our focus is on hotspots moving along two distinct types of trajectories: plunging geodesics and homoclinic orbits \cite{Levin:2008yp}. We analyze the time-averaged polarized images of these hotspots, carefully examining variations in total intensity, the polarization vector along the trajectory, and the Stokes parameters under different magnetic field configurations. We identify a range of polarization characteristics associated with different trajectories. Furthermore, we apply this methodology to study the polarized image of the plunging region within an accretion disk, exploring the distributions of total intensity and polarization vectors across the disk plane.

The remaining parts of the paper are organized as follows. In Sec. \ref{theKerr}, we review the timelike geodesic equations in the equatorial plane of the Kerr spacetime, and discuss the characteristics of plunging and homoclinic orbits. In Sec. \ref{imaginghs}, we discuss the light propagations, the imaging screen for the on-axis observer, gravitational lensing formulas, and synchrotron radiation at the light source. In Sec. \ref{results}, we showcase the polarization patterns of different trajectories and discuss their distinctive features. We conclude our study in Sec. \ref{sum}. We will work in the unit with $G=c=1$.

\section{Trajectory of the hotspot}\label{theKerr}
We begin with a brief review of the trajectory of the hotspot on the equatorial plane in a Kerr spacetime. In the Boyer-Lindquist coordinates, the line element of the spacetime takes the form
\begin{align}
\mathrm{d}s^2=&-\left(1-\frac{2Mr}{\Sigma}\right)\mathrm{d}t^2+\frac{\Sigma}{(r-r_+)(r-r_-)}\mathrm{d}r^2+\Sigma\mathrm{d}\theta^2-\frac{4Mra}{\Sigma}\mathrm{sin}^2\theta\mathrm{d}t\mathrm{d}\phi\nn \\ & + \sin^2{\theta}\left(r^2+a^2+\frac{2Mra^2}{\Sigma}\mathrm{sin}^2\theta\right)\mathrm{d}\phi^2  \,,
\end{align}
where $ \Sigma=r^2+a^2\cos^2\theta$, $M$ and $a$ are the ADM mass and spin parameters, respectively. The event horizons reside at $r_{\pm}=M\pm M\sqrt{1-a^2}$. In subsequent discussions we focus on the case $a > 0$.

\subsection{Timelike circular orbits}

The four-velocity vectors are represented as $U^{\mu}=\mathrm{d}x^{\mu}/\mathrm{d}\tau$ with $\tau$ the proper time. The Killing symmetries along $t$ and $\phi$ directions yield conserved energy density and angular momentum density, symbolized by $E \equiv -U_t$ and $L \equiv U_{\phi}$, respectively. The geodesic equation in the equatorial plane reads
\bea
(U^{r})^{2}+\f{2M}{r} - \f{L^2-a^2(E^2-1)}{r^2} + \f{2M(L-aE)^2}{r^3} =E^2 -1 \,.
\label{(U^{r})^{2}+} 
\eea
For our purposes, we rewrite the above equation as $(U^{r})^{2}+V_{\rm eff}(r)=0$, where $V_{\rm eff}(r)$ functions as the effective potential,
\bea\label{Vr123}
V_{\rm eff}(r)=(E^2-1)\left(\frac{r_{1}}{r}-1\right)\left(\frac{r_{2}}{r}-1\right)\left(\frac{r_{3}}{r}-1\right) \,.
\eea
Here $r_{1}$, $r_{2}$ and $r_{3}$ represent the roots of $V_{\rm eff}(r)$. Typically, these roots are complex, suggesting that there may be zero to three physical roots outside the black hole, depending on the values of $E,L$ as well as the spin. The potential simplifies in the presence of a real double root, denoted as $r_1 = r_2 = r_c$. In this case, one has $V_{\rm eff}(r_{c})=V_{\rm eff}'(r_{c})=0$, indicating that the potential permits a circular orbit at $r=r_{c}$. The expressions for the conserved quantities of a circular orbit are provided by \cite{chandrasekhar1998mathematical}
\bea\label{cirEL}
L_c = \pm \frac{(Mr_c)^{1/2}}{\mathcal{D}}\left(1+ \frac {a^2}{r_c^2}\mp\frac{2aM^{1/2}}{r_c^{3/2}}\right)\,,\quad E_c  = \frac{1}{\mathcal{D}}\left(1-\frac {2M}{r_c}\pm\frac{aM^{1/2}}{r_c^{3/2}}\right)\,,
\eea
where
\bea
\mathcal{D}^2=1-\frac{3M}{r_c}\pm\frac{2aM^{1/2}}{r_c^{3/2}} \, ,
\eea
and the ``$\pm$'' indicates the prograde/retrograde orbits. Comparing Eq.~\eqref{(U^{r})^{2}+} and Eq.~\eqref{Vr123}, we get the expression of $r_3$ for the case with double root,
\bea\label{r3}
	r_3 =\frac{2M}{r^2_c}\frac{(L_c-aE_c)^2}{1-E_c^2} \,.
\eea
Hence, for a potential that possesses a double root at $r=r_c$, the conserved quantities and the third root are dictated by the value of $r_c$. The stability of circular orbits is determined by the sign of $V''_{\rm eff}(r_{c})$. Marginal stability occurs when $V''_{\rm eff}(r_{c}) = 0$, which yields the radius of the innermost-stable-circular orbit (ISCO), denoted by $r_I$. The circular orbits outside the ISCO are stable and converge towards their Newtonian behavior at substantial radii. For $r_c < r_I$, the circular orbit becomes unstable under perturbations. The radius of the ISCO is given by 
\bea
&&r_I/M=3+Z_2 \mp \sqrt{(3-Z_1)(3+Z_1+2Z_2)}\,,\nn \\
&&Z_1=1+(1-a_\star^2)^{1/3}\left[(1+a_\star)^{1/3}+(1-a_\star)^{1/3}\right]\,, \quad Z_2=\sqrt{3a_\star^2+Z_1^2} \, ,
\eea
where $a_\star=a / M$ is the dimensionless spin parameter. In addition to the stability of circular orbits, another characteristic pertains to whether the orbits are bound or unbound. The notion of the bound orbits can be understood from Eq.~\eqref{r3}: the root $r_3$ exists only when $E_c < 1$. This leads to a class of pseudo-circular orbits where a particle, once perturbed outward from $r = r_c$, travels out to $r_3$ and subsequently returns to $r_c$. For $E_c \geq 1$, $r_3$ ceases to exist, and a point particle at the circular orbit $r = r_c$ will become unbound and escape to infinity under perturbations. The marginal case is the innermost-bound-circular orbit (IBCO), the radius of which is determined by setting $E_c = 1$ in Eq.~\eqref{cirEL}, resulting in 
\bea
r_{ib}/M=\left(\, 1+\sqrt{1\mp a_\star}\,\right)^2\,.
\eea
For $r_{ib} < r_c < r_I$, the circular orbits have $E_c < 1$ and remain bound, though they are unstable. The orbits inside the IBCO has $E_c > 1$ and are both unstable and unbound.

\subsection{Inspiraling orbits and homoclinic orbits}

\begin{figure}[ht!]
	\centering
	\includegraphics[width=5.2in]{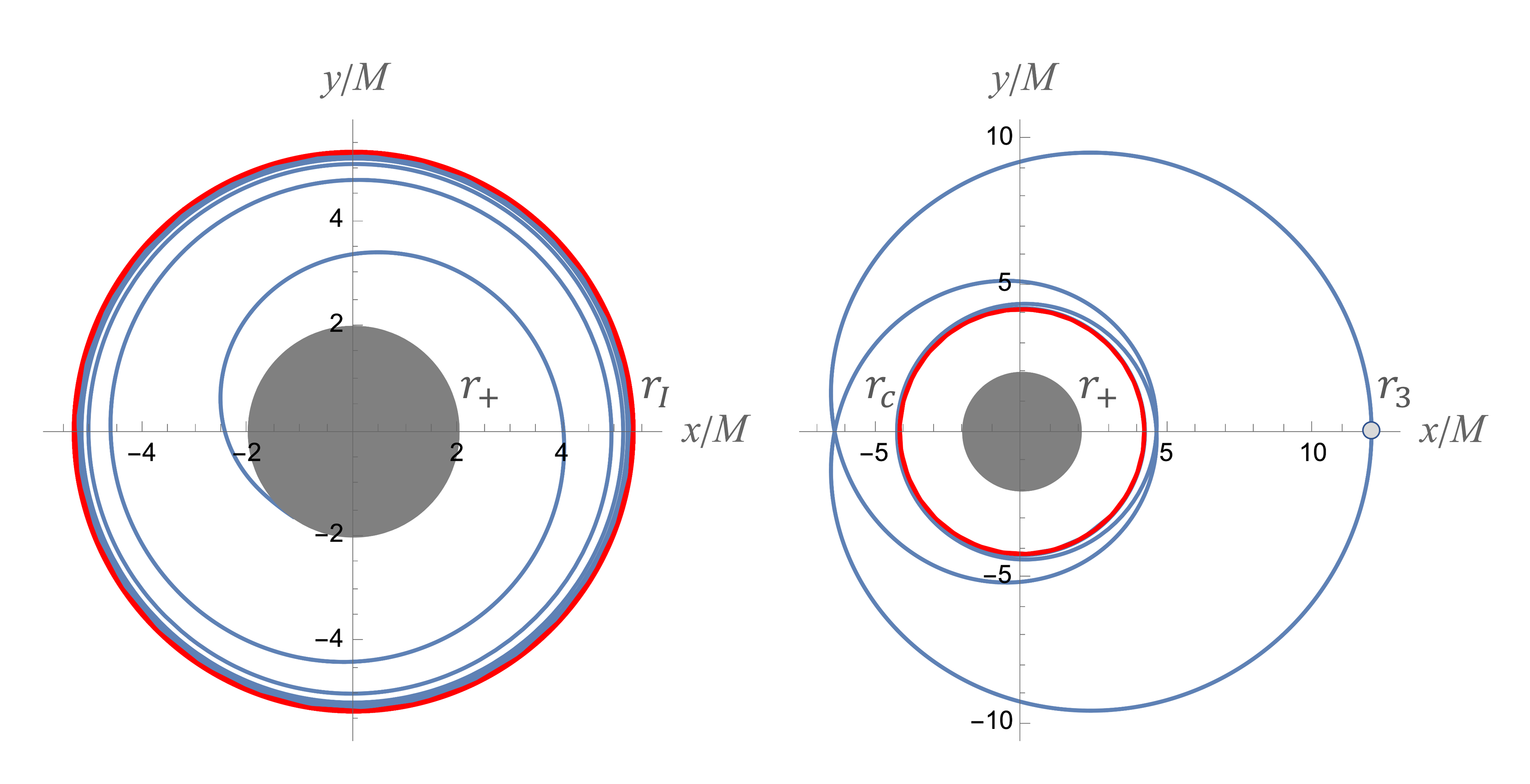}
	\centering
	\caption{Left: The red circle represents the prograde ISCO, $r = r_I$, and the blue curve represents the inspiral orbit from the ISCO. Right: The red circle represents the prograde, bound circular orbit $r_c = r_{ib} + 0.6 M\,$, and the blue curve represents the homoclinic orbit outward from $r_c$. The gray dot denotes the radius of $r_3$. The black hole spin is set to be $a = 0.2M$.}
	\label{Orbs}
\end{figure}

For an unstable circular orbit, the hotspot will deviate from the circle under small perturbations. Consider a hotspot with $E_c = E(r_c), L_c = L(r_c)$, which is perturbed off the orbit $r = r_c$. If $r_I \geq r_c > r_{ib}$, the equations of motion are
\bea
\frac{\md r}{\md\tau}=  \pm_r \left( \f{r_c}{r}-1 \right) \sqrt{\f{2M(L_c-aE_c)^2}{r_c^2 \, r}+E_c^2-1}\, ,  \quad   \frac{\md\p}{\md\tau}= \frac{2MaE_c+L_c(r-2M)}{r(r-r_+)(r-r_-)}\, ,
\label{azimuthal}
\eea
from which we may obtain the trajectory of the hotspot parameterized by $r$,
\bea\label{integral}
\sqrt{1-E_c^2}\,\p=\int\frac{r^{3/2}\left[2MaE_c /r+L_c(1-2M/r)\right]}{(r-r_+)(r-r_-)(r-r_c)\sqrt{r_3-r}}\,\mathrm{d}r \,.
\eea
It is evident that there are poles at $r_{\pm}$ and $r_c$ in the integral, suggesting that the hotspot warps infinitely near the poles and approaches the radius in the asymptotic future or past. The poles at $r_{\pm}$ correspond to the frame-dragging effect of the Kerr black hole, while the pole at $r_c$ reveals the nature of the unstable circular orbit. In the regime $|r - r_c| \ll r_c$, if the hotspot moves from $r_1$ to $r_2$ within a time interval $\D\tau = \tau_2 - \tau_1$, one finds
\bea\label{Lya}
\f{r_2-r_c}{r_1-r_c} \approx \, e^{\pm\g (\tau_2-\tau_1)} \, ,\quad  \g = \f{1}{r_c}\sqrt{\f{2M}{r_c^3}(L_c-aE_c)^2 + E_c^2 -1}  \, ,
\eea
where the ``$ \pm$'' denotes whether the motion is outward/inward relative to $r=r_c$. The parameter $\g$ is the Lyapunov exponent encoding the asymptotic behavior near $r = r_c$ \cite{Levin:2008yp}. At the ISCO, we have $V_{\rm eff}(r_{I})=V'_{\rm eff}(r_{I})=V''_{\rm eff}(r_{I})=0$, resulting in a triple root, $r_c = r_3 = r_I$.  Under small perturbations, the hotspot at the ISCO will plunge into the horizon, whose radial velocity takes a simple form \cite{chandrasekhar1998mathematical}
\bea\label{urisco}
U_{I}^{r}=-\sqrt{\frac{2M}{3r_{I}}}\left(\frac{r_{I}}{r}-1\right)^{3/2}.
\eea
In this case, Eq.~\eqref{integral} can be integrated to yield the trajectory function parameterized by $r$ \cite{Mummery:2022ana,Ko:2023igf}
\bea\label{f012isco}
&&\phi(r)=f_+(r)+f_-(r)+f_I(r)\,, \nn \\
&&f_{\pm}(r)=\pm\left[ \frac{2L_c r_{\pm}^{3/2}+4Mr_{\pm}^{1/2}(aE_c-L_c)}{(r_+-r_-)(r_c-r_\pm)\sqrt{r_3-r_\pm}}\right]
\frac{1}{\sqrt{1-E_c^2}}\tanh^{-1}\left( \sqrt{\frac{r_3/r-1}{r_3/r_\pm-1}}\right)\,, \nn \\
&&f_I(r)=\sqrt{\frac{6r_I}{M}}\left( \frac{2M(L_c-aE_c)-r_IL_c}{r_I^2-2Mr_I+a^2}\right)\sqrt{\frac{r}{r_I-r}} \, .
\eea
The trajectories with different initial conditions can be related to Eq.~\eqref{f012isco} through the rotations along the $\p$ direction. From Eq.~\eqref{f012isco}, we see that $f_{\pm}(r)$ governs the asymptotic behavior near $r_{\pm}$ respectively, while $f_I(r)$ characterizes the asymptotic behavior near the ISCO, which differs from the exponential behavior near $r_c$ given by Eq.~\eqref{Lya}. The left panel of Fig. \ref{Orbs} showcases an inspiral orbit, from which it is clear that the trajectory near $r_I$ exhibits significantly denser loops.

For a circular orbit inside the ISCO, i.e., $r_{ib} < r_c < r_I$, the orbit is unstable but remains bound. There emerge two types of trajectories under the perturbations at $r = r_c$: the orbits  spiral outwards from $r_c$ towards $r_3$, or the orbits spiral inwards from $r_c$ and then plunge into the outer horizon. The latter case is qualitatively similar to the inspiral orbit from the ISCO. For the orbits in the former case, the trajectory function parameterized by $r$ takes the form of
\bea\label{f012homo}
&&\phi(r)=f_+(r)+f_-(r)+f_h(r)\,, \nn \\
&& f_{h}(r)=-\left[ \frac{2L_cr_{c}^{3/2}+4Mr_{c}^{1/2}(aE_c-L_c)}{(r_c-r_+)(r_c-r_-)\sqrt{r_3-r_c}}\right]
\frac{1}{\sqrt{1-E_c^2}}\tanh^{-1}\left( \sqrt{\frac{r_3/r-1}{r_3/r_c-1}}\right)\,,
\eea
where $f_{\pm}(r)$ is the same as in Eq.~\eqref{f012isco}. These orbits are referred to as homoclinic orbits \cite{Levin:2008yp,Li:2023bgn}. Physically, a homoclinic orbit corresponds to a hotspot forming at an unstable but bound circular orbit, and is subsequently perturbed away from the circular orbit. The right panel in Fig. \ref{Orbs} depicts a homoclinic orbit with parameters $a = 0.2, r_c = r_{ib} + 0.6 M$, and $r_3 \approx 11.96$. In this scenario, the hotspot is perturbed outwards from $r = r_c$, circles around it multiple times, and eventually reaches the ``aphelion" at $r = r_3$. Subsequently, the hotspot retreats from the aphelion back towards $r = r_c$.

\section{Imaging the hotspot}\label{imaginghs}

As our interest lies in the images of the hotspot, we need to compute the light propagations from the hotspot to a distant observer. In this section, we initially revisit the basic equations of light propagation in Kerr spacetime. Then, we discuss the map between the screen coordinates of the observer and the hotspot's position, and provide a new fitting function of the map, which proves crucial for an analytical examination of the hotspot's images. Subsequently, we introduce the setting for synchrotron radiation of the hotspot under external magnetic field.

\subsection{Light propagations}
It is well-established that in a Kerr spacetime, a photon has three conserved quantities along its trajectory, namely the energy, angular momentum, and the Carter constant, expressed as
\bea
\mathcal{E}=-p_{t},\quad \mathcal{L}=p_{\phi},\quad \mathcal{Q}=p^2_{\theta}-\cos^2\theta(a^{2}p_{t}^{2}-p_{\phi}^{2}\csc^2\theta). \label{conserved quantities}
\eea
These quantities originate from the Killing symmetries along $t, \p$ directions and a high-rank symmetry \cite{Carter:1968rr}. The magnitude of $\mathcal{E}$ can be absorbed into the scaling of the affine parameter and does not affect null geodesic trajectories. Therefore, we work with energy-rescaled quantities, defined as
\bea
\lambda=\frac{\mathcal{L}}{\mathcal{E}}\,,\quad \eta=\frac{\mathcal{Q}}{\mathcal{E}^2} \, ,
\eea
also referred to as the impact parameters. A photon with an affinely parameterized trajectory has the energy-rescaled four-momentum
\bea
p_{\mu}\mathrm{d}x^{\mu}=-\mathrm{d}t
\pm_{r}\frac{\sqrt{\mathcal{R}(r)}}{(r-r_+)(r-r_-) }\mathrm{d}r
\pm_{\theta}\sqrt{\Theta(\theta)} \mathrm{d}\t
+\lambda\mathrm{d}\phi \, ,
\label{four-momentum}
\eea
given in terms of the radial and angular potentials
\bea
&&\mathcal{R}(r)=(r^2+a^2-a\lambda)^2-(r-r_+)(r-r_-) [\eta+(\lambda-a)^2]\,,\nn \\
&&\Theta(\theta)=\eta+a^2\cos^2\theta-\lambda^2\cot^2\theta.
\eea
The symbols $\pm_{r}$ and $\pm_{\theta}$ indicate the sign of $p_r$ and $p_\theta$, respectively. The variations of the impact parameters change the shape of the potentials and determine the behaviors of null geodesics. For our purpose, we restrict ourselves to the case of $\eta > 0$, corresponding to the light rays that can reach the equatorial plane.

As the hotspot emits synchrotron radiation under an external magnetic field, we need to consider the transport of the polarization vector along a light ray. The unit-norm spacelike polarized vector, denoted by $f^\m$, is orthogonal to the photon's four-momentum and parallel transported along the trajectory,
\bea
f \cdot p=0\,, \quad f\cdot f= 1\, , \quad p \cdot \nabla f^\n=0\,,
\eea
where we have omitted the indices when performing the inner product. 
With the help of the Newman-Penrose null tetrad $\{l, n, m, \bar{m}\}$, one can construct a conserved quantity for the polarization vector along null geodesics \cite{Walker:1970un}, typically written as 
\bea \label{PW}
&&\kappa \equiv \k_{1}+i \k_{2} = (A-iB)(r-ia\cos\t) \, , \\
&&A= (p^{t} f^{r} - p^r f^t) + a \sin^2{\t} (p^r f^\p-p^\p f^r) \, , \nn \\
&&B = [(r^2+a^2) (p^{\p} f^{\t} - p^{\t} f^{\p}) -a (p^t f^\t - p^\t f^t)]\sin{\t} \,,
\eea
where $\kappa_1, \kappa_2$ represent the real and imagine part of $\kappa$, respectively. The conserved quantity is often referred to as the Penrose-Walker constant.

\subsection{Coordinates on the screen}\label{screen} 

The observations indicate small inclination angles for the supermassive black holes \cite{EventHorizonTelescope:2019dse, GRAVITY:2023avo, EventHorizonTelescope:2024viii}. In this study, for simplicity, we restrict ourselves to an on-axis observer. From the angular potential, it is evident that only the photons with zero angular momentum, $\lm = 0$, can approach the spin axis. When a photon reaches the screen of the on-axis observer at the distance, its position of arrival is given by the polar coordinates \cite{Gelles:2021kti}
\bea\label{polarcoord}
\rho = \f{\sqrt{\eta +a^2}}{r_o}\, ,  \quad  \varphi=\phi_o \, ,
\eea
where $\rho$ represents the angular radius of the photon on the screen, $r_o$ and $\p_o$ are the radius and azimuthal angle of the arriving photon, respectively. In what follows we take a rescaling  
\bea
\rho \rightarrow \f{r_o}{M} \rho
\eea
to simplify the calculation. The polar coordinates encode the four-momentum of the photon and allows us to determine the route of the arriving photon. Our study focuses on a hotspot at the equatorial plane that serves as the sole light source. For the photons reaching $(\rho, \varphi)$, the emission point $(r_s, \t_s, \p_s)$ is determined by combining the expression of the photon's trajectory with the constrain $\t_s = \pi/2$, yielding a relationship between the hotspot's position and the arriving photon's polar coordinates, the latter of which provides the hotspot's image point on the observer's screen.

Owing to the strong gravitational lensing effect, a null geodesic can loop around the black hole multiple times, potentially resulting in more than one image point associated with a single equatorial plane position, known as the lensed images of the hotspot \cite{Huang:2024wpj}. However, our interest lies solely in the direct image of the hotspot, which is produced by the photons that do not loop around the black hole  and directly escape to the distant observer. For this direct image, the relationship between $(r_s, \p_s)$ and $(\rho,\varphi)$ can be well approximated by the analytical expansions presented in \cite{Gelles:2021kti}, 
\bea
&&\f{r_s}{M}=\rho -1+\frac{1-a^{2}}{2\rho}+\frac{3(5\pi-16)}{4\rho^2}+\mathcal{O}(1/\rho^3)\,, \label{inverse1}  \\
&&\rho=\f{r_s}{M}+1+\frac{a^2-M^2}{2Mr_s}+\frac{50M^2-2a^2-15M^2\pi}{4r_s^2}+\mathcal{O}(M^3/r_s^3)\,. \label{inverse2}
\eea
When deriving the above equations, the limit $r_s \gg M$ has been used. However, the above equations work quite well even at the horizon scale ($r_s \sim r_+$) for all Kerr spin, as demonstrated in the Fig.5 in \cite{Gralla:2019drh}. Therefore, we can safely use Eq.~\eqref{inverse1} and Eq.~\eqref{inverse2} to study the images of a hotspot close to the event horizon.

\begin{figure}[ht!]
	\centering
	\includegraphics[width=6.6in]{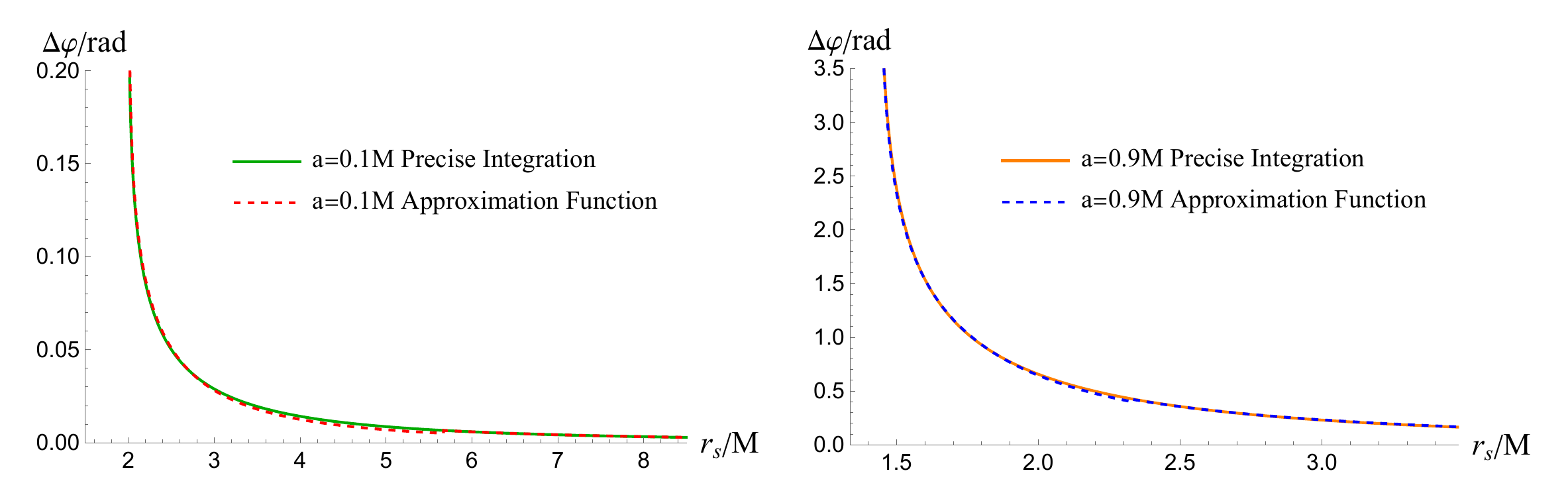}\\
	\includegraphics[width=3.5in]{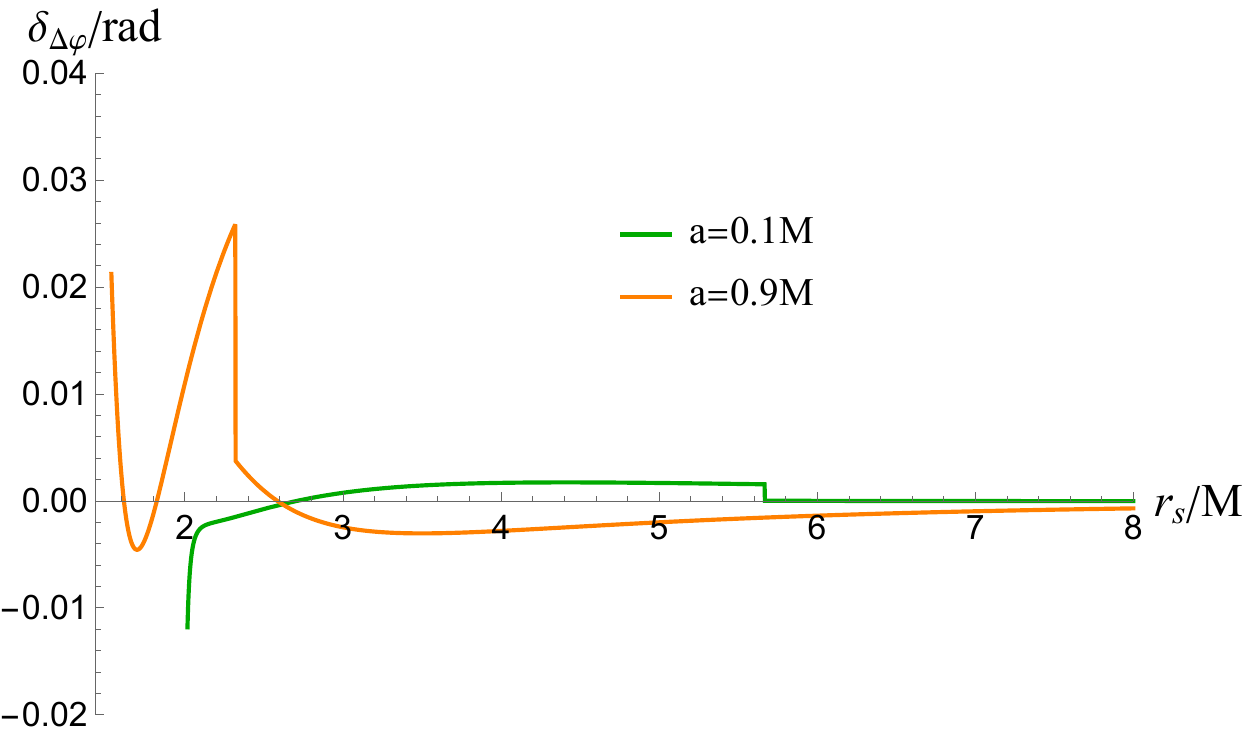}
	\centering
	\caption{Comparision between Eq.~\eqref{exactvarphi} and Eq.~\eqref{appphi} for two different Kerr parameters $a=0.1M$ and $a=0.9M$. In the upper plots, the solid lines represent the exact results, $\Delta\varphi_\text{int}$, obtained by numerically integrating Eq.\eqref{exactvarphi}, while the dashed lines correspond to the approximate results, $\Delta\varphi_\text{fit}$, computed using the fitted expression in Eq.\eqref{appphi}. The lower plot illustrates the deviations, defined as $\delta_{\Delta\varphi} \equiv \Delta\varphi_\text{int} - \Delta\varphi_\text{fit}$.}
	\label{Com}
\end{figure}

For an axisymmetric light source, such as a time-averaged accretion disk, Eq.~\eqref{inverse1} and Eq.~\eqref{inverse2} are sufficient to determine the image. This is because the emission profile does not depend on $\p_s$, and the image does not depend on $\varphi$. However, when the hotspot serves as a point light source, a significant distortion along the $\p$ direction occurs between the hotspot and its image on the screen, i.e., $|\varphi - \p_s|$ could be large. This distortion originates from the frame-dragging effect on the light trajectory. Consequently, it is necessary to determine the map $\varphi(r_s, \p_s)$ for hotspot imaging. For a photon emitted at $(r_s, \p_s)$, the angle of arrival takes
\bea\label{exactvarphi}
\varphi = \p_s +  \int_{r_s}^{\infty} \f{2Ma rdr}{(r-r_+)(r-r_-) \sqrt{\mathcal{R}}} \,, \quad \mathcal{R}=(r^2+a^2)^2-(r-r_+)(r-r_-)(\eta+a^2) \, ,
\eea
where $\eta$ is related to $r_s$ through Eq.~\eqref{polarcoord}  and Eq.~\eqref{inverse2}, for the direct image. To make a preliminary discussion, we take the limit $r_s \gg r_+$ in Eq.~\eqref{exactvarphi}, and the radial potential can be expanded to $R \approx r^4 - r_s^2 r^2 + \mO\left(  r_s r^2 \right) + \mO\left(  r_s^2 r \right) $. This expansion is valid if $r$ is greater than $r_s$, a condition easily met for the light rays associated with the direct image. If we consider only the first two terms in the expansion, the integral simplifies to
\bea\label{app1}
\Delta\varphi\equiv\varphi - \p_s  \approx  \int_{r_s}^{\infty} \f{2Ma dr}{ r^2 \sqrt{r^2-r_s^2}} =  \f{2Ma}{r_s^2} \, .
\eea
This expression works well for a large $r_s$. However, when $r \rightarrow r_+$, the integral in Eq.~\eqref{exactvarphi} diverges for a nonzero spin, a phenomenon attributed to the inevitable dragging effect, and Eq.~\eqref{app1} is not applicable in this region. To derive an approximate equation for $r_+ \leq r < \infty$, one has to take the pole at $r = r_+$ into account. The purely analytical expansion for Eq.~\eqref{exactvarphi} is  complicated, and, in this work we resort to numerical schemes for function fitting. Based on the asymptotic behavior described by Eq.~ \eqref{app1} and the behavior near $r = r_+$, we develop a piecewise function to approximate Eq.~\eqref{exactvarphi}, which takes the form
\bea\label{appphi}
\Delta\varphi\approx
\left\{
\begin{aligned}
	&\,\, \frac{2Ma}{r_s^2}\left[1-\frac{\td{\b}}{(\frac{r_s}{r_+}-1)^{\td{\a}}}\ln{(\frac{r_s}{r_+}-1)}\right] \,,  \quad \,\, r_s\textless r_I\, , \\
	&\,\, \frac{2Ma}{r_s^2}\left[1+\frac{\td{\g}}{(\frac{r_s}{r_+}-1)}\right] \, , \quad \,\,	r_s\geq r_I\, , \\
\end{aligned}
\right.
\eea
where $r_I$ is the prograde ISCO, and $\td{\a}, \td{\b}, \td{\g}$ are three functions of $a_{\star} = a/M$,
\bea
\left\{
\begin{aligned}
	\td{\a}&=0.035(1-a_{\star})+\frac{0.0059}{(1-a_{\star})^{0.4577}}+0.1163 \,, \\
	\td{\b}&=0.2093\arctan{(a_{\star}^{12})}+0.3467 \,, \\
	\td{\g}&=0.07815\arctan{(a_{\star}^3)}+0.0983\,.
\end{aligned}
\right.
\eea
Fortunately, the approximation function Eq.~\eqref{appphi} works well for a large parameter range of $a_{\star}$. In Fig. \ref{Com} we present the comparison between the approximation function Eq.~\eqref{appphi} and the precise result of Eq.~\eqref{exactvarphi}, for a small spin $a_{\star} = 0.1$ and a high spin $a_{\star} = 0.9$. We also plot the deviation from the precise result, $\delta_{\Delta\varphi}=\Delta\varphi_\text{int}-\Delta\varphi_\text{fit}$. The deviation function $\delta_{\Delta\varphi}$ is slightly discontinuous at the ISCO, where the largest error occurs. For $a = 0.9M$, the maximum error is $\delta_{\Delta\varphi\text{max}} < 0.03 \approx 1.72^\circ$, while for $a = 0.1M$, the maximum error is even smaller. Therefore, the error in using the approximation function is rather small and can be neglected in subsequent discussions. By using Eq.~\eqref{appphi}, we can easily obtain the hotspot images for the on-axis observer, without the need to deal with lengthy elliptic integrals.

\subsection{Synchrotron radiation of the hotspot}
In order to obtain the polarized image at the screen, we need to specify the synchrotron radiation of the hotspot. The polarized intensity emitted from the hotspot is determined by its magnetic field, $B_\m = -\star F_{\m\n} U^\n$, with $\star F_{\m\n}$ being the Hodge dual of the electromagnetic tensor, and $U^\m$ being the four-velocity of the hotspot. In the hotspot's comoving frame, we introduce the spatial four-momentum of the radiated photons, as well as the magnetic field perpendicular to this momentum, 
\bea
p_{\perp}^\m = p^\m + (U \cdot p)U^\m \, , \quad B^{\m}_{\perp} = B^\m - \f{B\cdot  p}{(U\cdot  p)^2} \, p_{\perp}^\m \,.
\eea
From the above expressions we find $U \cdot p_{\perp} = U \cdot B_{\perp} = p_{\perp} \cdot B_{\perp} = 0$. For thermal synchrotron emission, the polarization vector is perpendicular to the photon momentum and the magnetic field \cite{1979Lightman}. Thus, we can express the polarization vector as 
\bea
f^\m = -\f{\epsilon^{\m\n\a\b} U_\n p_\a (B_{\perp})_\b }{(U\cdot p)\sqrt{B_{\perp}^2} }\, .
\eea
Besides, the emitted total intensity concentrates on the direction perpendicular to the magnetic field in the hotspot's comoving frame. 
Following \cite{EventHorizonTelescope:2021btj}, we set the total intensity to be $\mI = \xi B_{\perp}^2$, with the overall coefficient $\xi$ determined by the details of the hotspot. In what follows we simply set $\xi = 1$.

Strictly speaking, the structure of the magnetic field within the hotspot depends on the specific mechanism that generates the hotspot. In the present study, we assume that the hotspot is embedded within a strong background magnetic field, whose structure remains unaffected by the motion of the hotspot. We consider two types of magnetic fields, with vertical and radial configurations, to study the imprint of magnetic field structure on the polarization images. First, we  employ the source-free Papapetrou-Wald solution \cite{Wald:1974np} as the vertical field, in which the gauge potential reads 
\be\label{Wald}
A^\m \pl_\m = a\mathcal{B} \, \pl_{t} + \f{\mathcal{B} }{2}\,\pl_{\p} \, ,
\ee
where the constant $\mathcal{B} $ indicates the field strength observed by a distant static observer. Such vertical configuration implies a dynamical important magnetic field near the equatorial plane, which has been suggested by some observations \cite{GRAVITY:2020hwn, GRAVITY:2023avo}. 

For the raidal configuration, we consider a magnetic field sourced by a background accretion flow infalling from the infinity \cite{Hou:2023bep}, with four-velocity $\bar{U}_\m dx^\m = -dt - \sqrt{2r(r^2+a^2)}(r-r_+)^{-1}(r-r_-)^{-1} dr$. By considering the ideal-MHD condition, the electromagnetic tensor near the equatorial plane takes
\bea\label{Radial}
F_{\t\p} = \Psi \, ,  ~~~~F_{r\t} =  \Psi\,\f{\bar{U}^\p}{\bar{U}^r}\, .
\eea
Here, $\Psi$ denotes the overall strength and keeps constant along the streamline. Within the accretion flow the magnetic field lines align with the streamlines, which appear as a radial pattern at a distance from the black hole, but near the horizon $r_+$, they become highly curled due to the frame-dragging effect.

\section{Polarization patterns}\label{results}

The polarization state of the observed photons is encoded in the Stokes parameters \cite{EventHorizonTelescope:2021bee, EventHorizonTelescope:2024vii, GRAVITY:2023avo}. Following the convention in \cite{EventHorizonTelescope:2021bee}, we denote the Stokes parameters as $\{\mI_o, \mQ_o, \mU_o, \mathcal{V}_o\}$, where $\mI_o$ is the observed total intensity, $\mQ_o$ and $\mU_o$ describe the linear polarization components along two perpendicular axes, and $\mathcal{V}_o$ represents the circular polarization \cite{1979Lightman}. In the case of pure linear polarization, the Stokes parameters read
\bea
\mP_o =\mQ_o +  i \, \mU_o  = \mI_o \, e^{2i\chi_o}  \, , \quad \mathcal{V}_o = 0 \, ,
\eea
where $\mI_o$ is the observed total intensity, related to the emitted intensity through $\mI_o = g^4\mI$ with $g$ the redshift factor from the hotspot to the observer \cite{Lindquist:1966igj}. The electric vector polarization angle (EVPA), denoted by $\chi_o$, is obtained by projecting the polarization vector on the screen's tetrad. The result takes \cite{Gelles:2021kti}
\bea
\chi_o = \arctan{\f{\k_1}{\k_2 }} + \f{\pi}{2}+ \varphi = \chi_{net} + \varphi  \, ,
\eea
where we have introduced the ``net EVPA", $\chi_{net} = \chi_o - \varphi\,$, to denote the angle between the polarization vector and the line from the origin of the screen to the image point $(\rho, \varphi)$. The values of the EVPA and net EVPA are set to be $\chi_o, \chi_{net} \in [\,0, \pi]$, for the reason that only the direction of the linear polarization is meaningful. It becomes evident that it is $\chi_{net}$ that encodes the pattern of the polarization. For instance, for an axisymmetric light source, $\chi_{net} = 0, \pi$ reflects a radial pattern, with the polarization vectors all pointing to the $\rho$ direction; $\chi_{net} = \pi/2$ indicates a toroidal pattern, where the polarization vectors all point to the $\varphi$ direction. Moreover, the case of $ \pi/2 > \chi_{net} > 0$ corresponds to the ``counterclockwise" pattern, that is, relative to the radial pattern, the polarization vector is combed in the counterclockwise direction. On the contrary, $\pi > \chi_{net} > \pi/2 $ corresponds to the ``clockwise" pattern, with the polarization vector aligned clockwise.

\subsection{Results for inspiraling hotspots}

\begin{figure}[th!]
	\centering
	\includegraphics[width=6.6in]{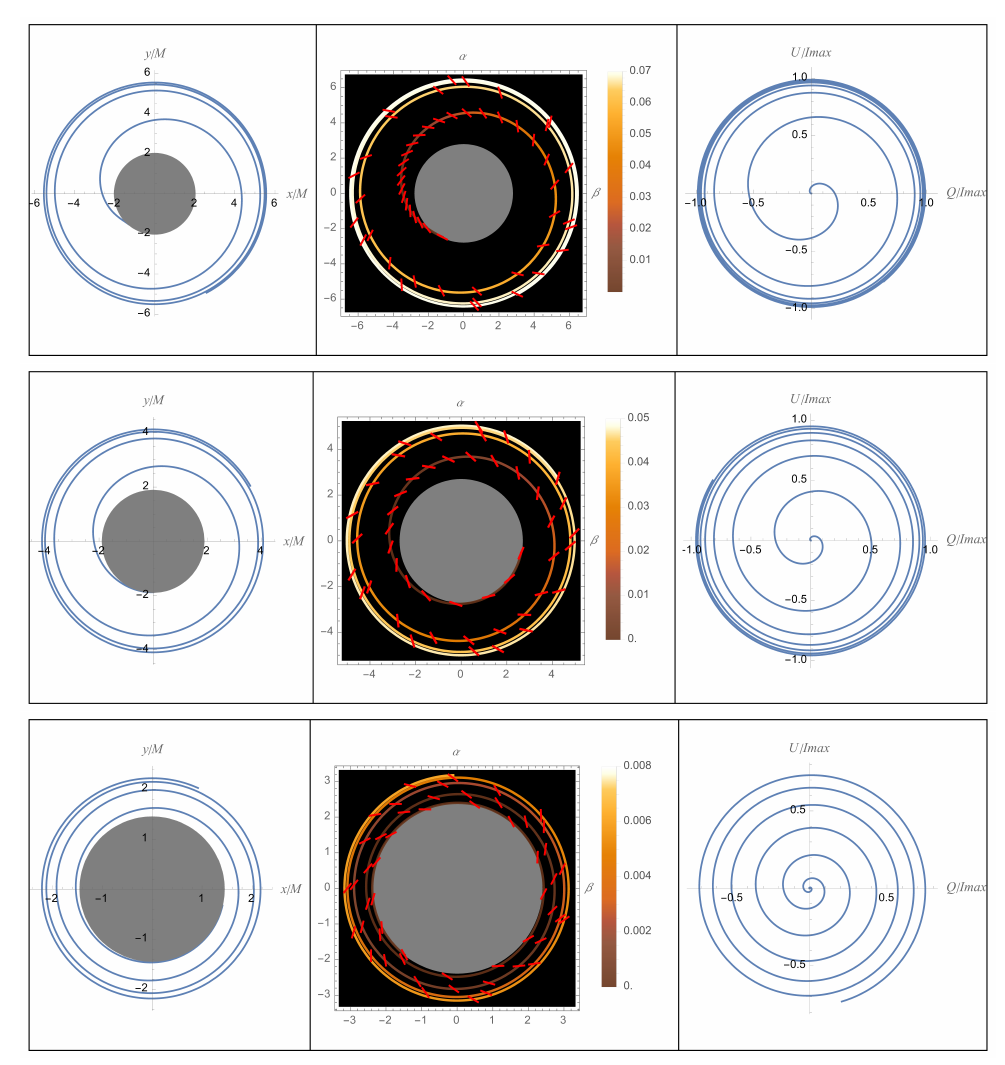}
	\centering
	\caption{From left to right, the columns present the trajectories, time-integrated polarized images, and the Q-U curve of the hotspot plunging from the prograde ISCO, under the vertical magnetic field Eq.~\eqref{Wald}. The constant $\mathcal{B}$ within the Papapetrou-Wald potential is set to be 1, without loss of generality. From top to bottom row, the black hole spin is set to be $a = 0.1M, 0.5M, 0.9M$.  For each case, the gray area denotes the projection of the black hole; the color bar represents the value of the total intensity; the polarization vector is represented by the red line segments on the screen; the Stokes parameters $Q, U$ are normalized by the maximal total intensity $I_{max}$ along the trajectory.}
	\label{inspWald}
\end{figure}

\begin{figure}[th!]
	\centering
	\includegraphics[width=6.6in]{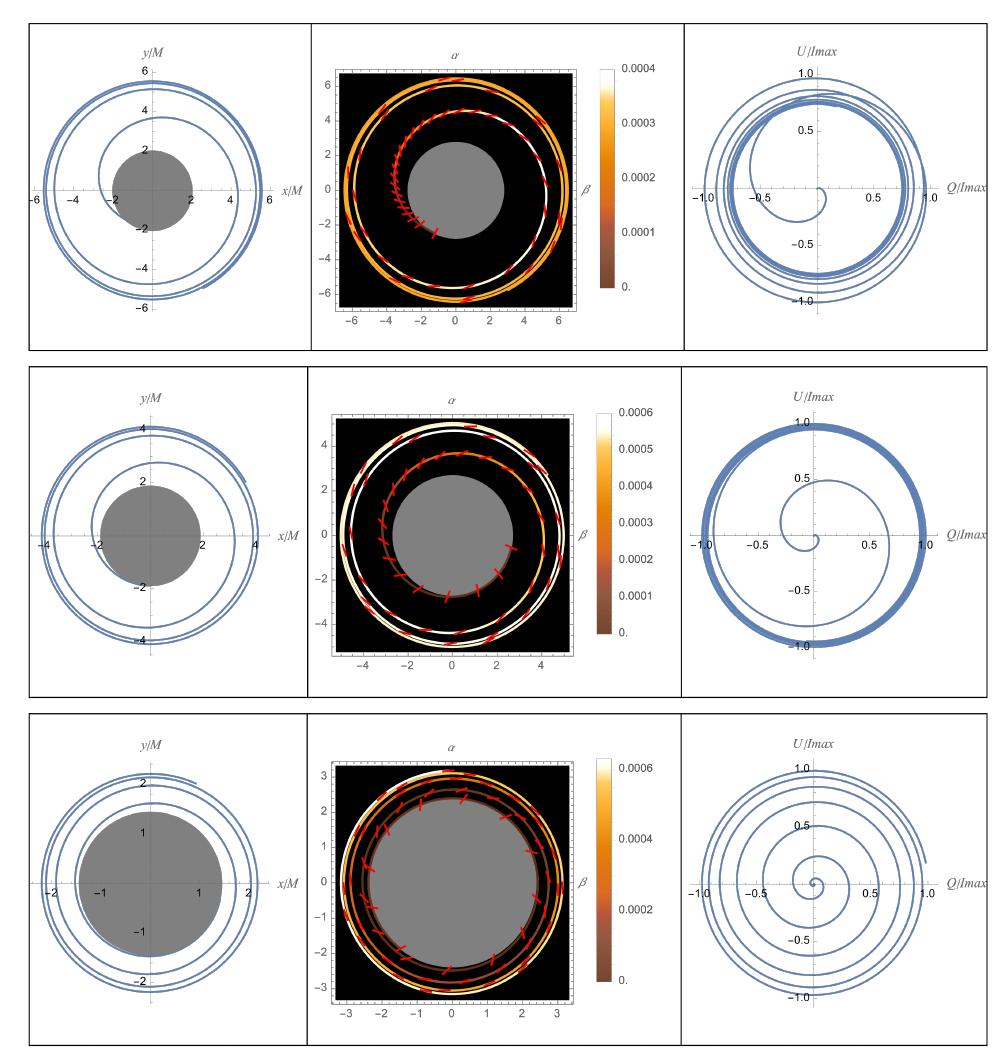}
	\centering
	\caption{From left to right, the columns present the trajectories, time-integrated polarized images, and the Q-U curve of the hotspot plunging from the prograde ISCO, under the radial magnetic field Eq.~\eqref{Radial}, where the constant $\Psi$ is set to be 1, without loss of generality. From top to bottom, the rows show the black hole spin set to $a = 0.1M, 0.5M,$ and $0.9M$.  For each case, the gray area denotes the projection of the black hole; the color bar represents the value of the total intensity; the polarization vector is represented by the red line segments on the screen; the Stokes parameters $Q, U$ are normalized by the maximal total intensity $I_{max}$ along the trajectory.}
	\label{inspRadial}
\end{figure}

In Fig. \ref{inspWald} we depict the trajectory and the time-integrated polarized image of the inspiraling hotspot under the vertical magnetic field Eq.~\eqref{Wald}, for various black hole spins. When depicting the images, we employ the Cartesian coordinates on the screen, $\a = \rho \cos{\varphi}\,, \b = \rho \sin{\varphi}$. Due to the gravitational redshift, the total intensity exhibits a decreasing trend as the hotspot approaches the horizon. In the right colume of Fig. \ref{inspWald}, the Stokes parameters along the hotspot's trajectory is displayed on the Q-U plane, which we call the Q-U curve. Similar to the trajectory, the Q-U curve exhibits a right-handed spiral shape\footnote{The chirality mentioned in this study is defined opposite to the direction of the observer's line of sight.}. From their definitions, $\mQ_o +  i \, \mU_o  = \mI_o \, e^{2i\chi_{net} + 2i\varphi}$, it is evident that the phase variation of $Q,U$ is mainly determined by the hotspot's motion in the $\p$ direction. 

Under the radial magnetic field Eq.~\eqref{Radial}, the polarization of the inspiraling hotspot significantly differs, as depicted in the middle column of Fig. \ref{inspRadial}. In particular, for the high-spin case, the polarization vector is 	nearly parallel to the trajectory near the ISCO and exhibits a clockwise pattern close to the horizon. From the right column of Fig. \ref{inspRadial}, we see that the Q-U curve first spirals outward and then back towards the origin for $a = 0.1M, 0.5M$, indicating a non-monotonic behavior of the total intensity along the trajectory.

\begin{figure}[th!]
	\centering
	\includegraphics[width=6.2in]{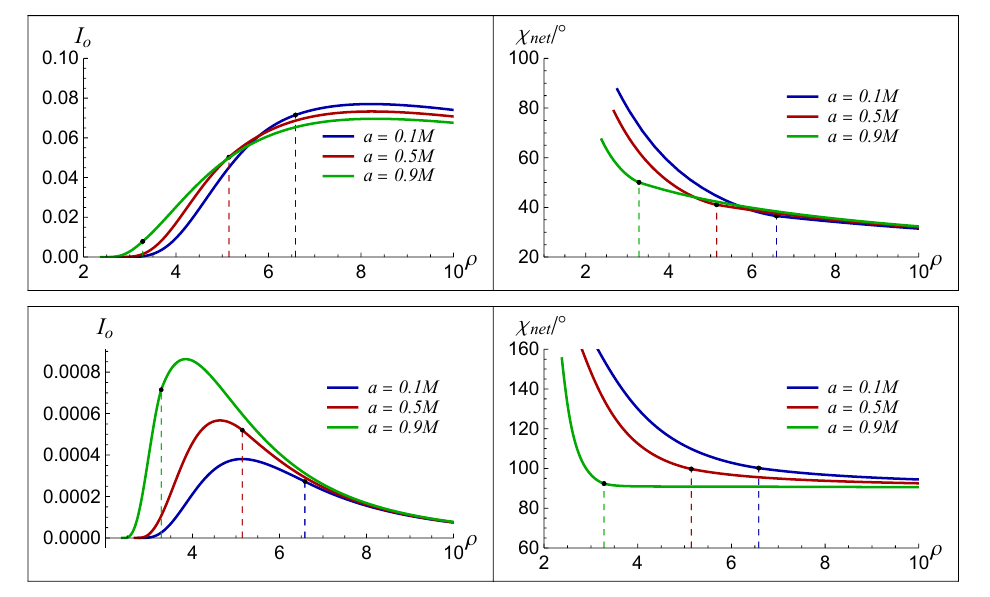}
	\centering
	\caption{The observed total intensity ($\mathcal{I}_o$) and net EVPA ($\chi_{net}$) along a prograde, inspiraling trajectory, under various black hole spin. The top and bottom rows represent the case of vertical and radial magnetic field, respectively. For each case, the dashed line represents the projected prograde ISCO on the observer's screen. To make a comparison, the region outside the ISCO is filled with luminous hotspots moving along prograde circular orbits. Since the constants for the two types of magnetic field ($\mathcal{B}$ and $\Psi$) only affect the overall magnitude of $\mathcal{I}_o$, we have normalized them to 1.}
	\label{intEVPA}
\end{figure}

The detailed behaviors of the total intensity and the net EVPA are presented in Fig. \ref{intEVPA}. The region outside the ISCO is set to be filled with luminous hotspots moving along prograde circular geodesics, which have been well studied in the literature. It is easy to see the distinction in $I_o$ and $\chi_{net}$ outside and inside the ISCO. Note that in all four subplots, the connections at the ISCO are continuous. Under the vertical magnetic field, the net EVPA increases as the hotspot nears the horizon. For a higher spin, $\chi_{net}$ displays a larger value at the prograde ISCO, but a smaller value at the horizon. However, the growth rate of  $\chi_{net}$ is largely insensitive to the spin, and in all cases the net EVPA exhibits a counterclockwise pattern, $\chi_{net} < \pi/2$, as shown in the top-right panel of Fig. \ref{intEVPA}. 

In the case of radial magnetic fields, the total intensity initially increases, and then drops towards zero as the hotspot approaches the horizon. This behavior can be attributed to the combined effects of gravitational redshift and the radial magnetic field. As the black hole continuously focuses the streamlines and magnetic field lines of the magneofluids, the magnetic field strength at smaller radii intensifies. When the hotspot spirals inward from the ISCO, the amplification of the magnetic field predominates in controlling its synchrotron radiation, which results in an increase in total intensity. However, as the hotspot gets closer to the black hole, the overpowering gravitational redshift reduces the radiations. In terms of the polarization pattern, the net EVPA maintains a fixed value of $\chi_{net} = \pi/2$ far from the black hole. This is due to the polarization vector being almost perpendicular to the radial magnetic field lines when $\rho \gg 1$. As $\rho$ decreases, $\chi_{net}$ varies slowly at first, then increases rapidly as the ISCO is crossed, consistently indicating a clockwise pattern.

\subsection{Results for hotspots from retrograde ISCO}

\begin{figure}[th!]
	\centering
	\includegraphics[width=6.6in]{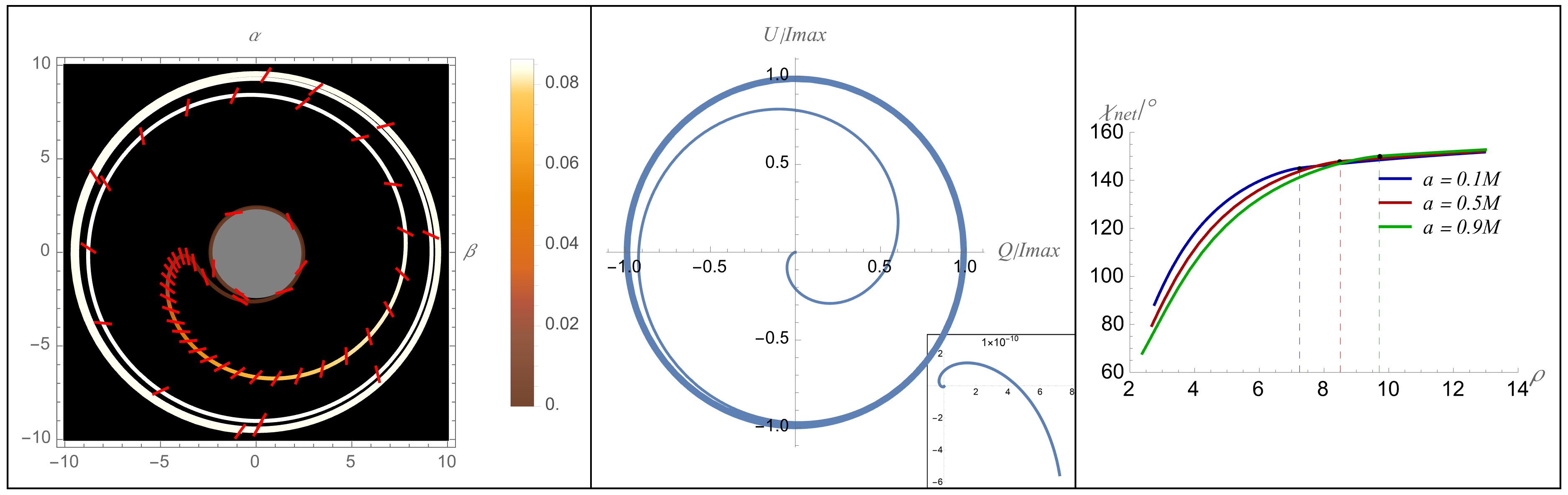}
	\centering
	\caption{The time-integrated polarized image, the Q-U curve, and the net EVPA of the hotspot plunging from the retrograde ISCO, under the vertical magnetic field Eq.~\eqref{Wald}. The constant $\mathcal{B}$ within the Papapetrou-Wald potential is set to be 1. In the left two panels, the black hole spin is set to be $a = 0.9M$. In the middle panel, we provide an enlarged view near the origin, where the scale of the coordinates should be multiplied by $10^{-10}$. In the right panel, the dashed line represents the projected retrograde ISCO on the observer's screen, outside which $\chi_{net}$ is produced by the hotspots moving along retrograde circular orbits.}
	\label{retro}
\end{figure}

Fig. \ref{retro} depicts the polarization pattern of a hotspot inspiraling from the retrograde ISCO, located at a greater distance from the black hole. Generally, under the vertical magnetic field, the net EVPA exhibits a clockwise pattern, $\pi>\chi_{net}>\pi/2$, as depicted in the right panel of Fig. \ref{retro}. As the hotspot approaches the black hole, it reaches a turning point where the rotation switches to prograde due to the frame-dragging effect, as demonstrated in the left panel of Fig. \ref{retro}. Near this turning point, the net EVPA reaches $\pi/2$ and then transitions to a counterclockwise pattern with $\pi/2 >\chi_{net}> 0$. Interestingly,  the Q-U curve in this case follows a left-handed spiral shape, contrary to the case that the hotspot spirals inward from the prograde ISCO. This behavior is influenced by the concentration of total intensity near the retrograde ISCO, which shapes the Q-U curve. In the magnified view in the lower right corner of the Q-U plane, a right-handed shape only emerges very close to the origin, corresponding to the polarization near the horizon.

\subsection{Results for homoclinic hotspots}

\begin{figure}[th!]
	\centering
	\includegraphics[width=6.6in]{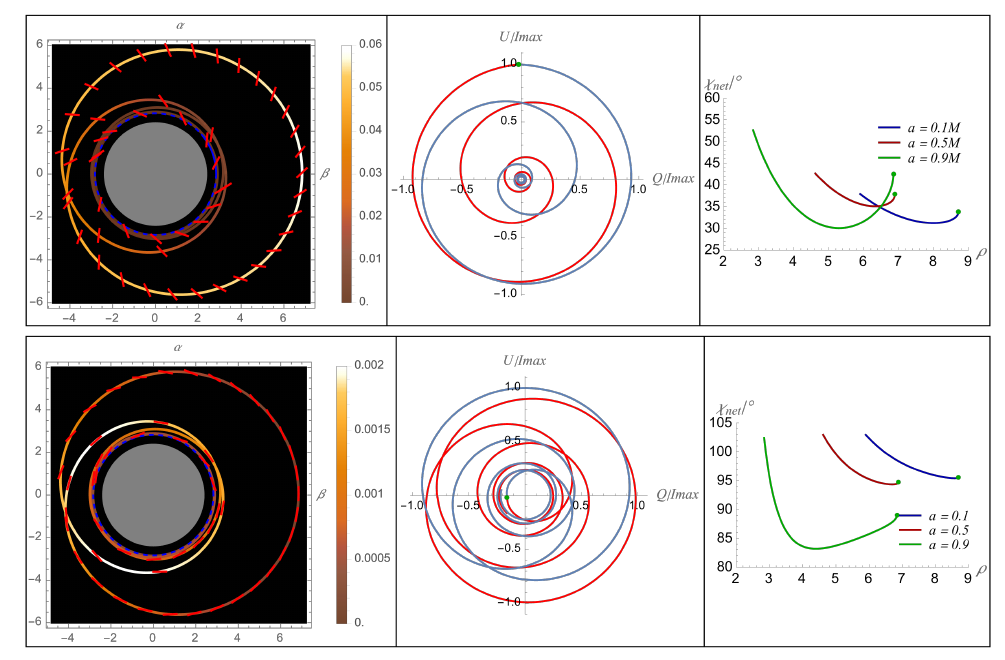}
	\centering
	\caption{The time-integrated polarized images, the Q-U curves, and the net EVPAs of the hotspot traveling along a homoclinic orbit, under the vertical magnetic field (top row) and radial magnetic field (bottom row). In the left two columns, we set $a=0.9M, r_c=r_{ib}+0.15M$. For each case, the projected $r_c$ is denoted by the blue dashed circle on the screen; the blue segment of the Q-U curve corresponds to the portion of the homoclinic orbit moving outward from $r_c$ to $r_3$, while the red segment represents the portion of the orbit falling from $r_3$ to $r_c$. The right column presents the variation of the net EVPA of the hotspots falling from $r_3$ (the green point) to $r_c$, under different spins. For the ease of presentation, the radii of $r_c$ are chosen as $r_c = r_{ib} + 1.2M/0.8M/0.15M$ for $a = 0.1M/0.5M/0.9M$.}
	\label{homo}
\end{figure}

Fig. \ref{homo} illustrates the polarization patterns of a hotspot along a homoclinic orbit, where the hotspot is perturbed to move outward from an unstable circular orbit. The net EVPA is greater under the radial magnetic field compared to the vertical magnetic field, a characteristic also observed in the case of an inspiraling hotspot. However, the Q-U curve exhibits a distinctive multi-loop structure, as illustrated in the middle column of Fig. \ref{homo}. In the previous studies, the Q-U loop has been considered to be a significant characteristic in the hotspot astrometry, attributed to the nonzero observation angle \cite{Vincent:2023sbw}. We now highlight that even when being observed on-axis, such a loop structure can arise from a homoclinic orbit. 

It is crucial to note that one-loop structure in the Q-U curves is a general feature of elliptical-like orbits, depending solely on the variation of total intensity along the orbit and the completion of a full period in $\varphi$. However, only the homoclinic orbit can generate the multi-loop structure on the Q-U plane in Fig. \ref{homo}, where the hotspot wraps around $r_c$ infinite times. Furthermore, the black-hole spinning introduces a slight breaking of mirror symmetry in the homoclinic orbit \cite{Levin:2008yp}. This asymmetry is amplified in the corresponding Q-U curve, as can be seen in the right column of Fig. \ref{homo}. Also, we want to stress that the hotspot itself may undergo evolution, causing variations in its emission profile. This may lead to the Q-U pattern evolving between subsequent orbits of the hotspot on a circular orbit. The difference in the loop structures in the Q-U plane caused by the homoclinic orbit and by the evolution of the hotspot  is an interesting issue worthy of further investigation.

\subsection{Effect of non-zero inclination angle}

In this section, we give a preliminary study on the effect of non-zero inclination angle on the polarization pattern of a hotspot. Since the imaging formula in Sec.\ref{screen} is not applicable for inclined observers, we employ the numerical backward ray-tracing method, as described in \cite{Zhang:2023cuw, Huang:2024wpj}. Besides, we use the Cartesian coordinates $(\alpha, \beta)$ introduced in \cite{Bardeen:1973tla} to cover the screen.

\begin{figure}[th!]
	\centering
	\includegraphics[width=6.2in]{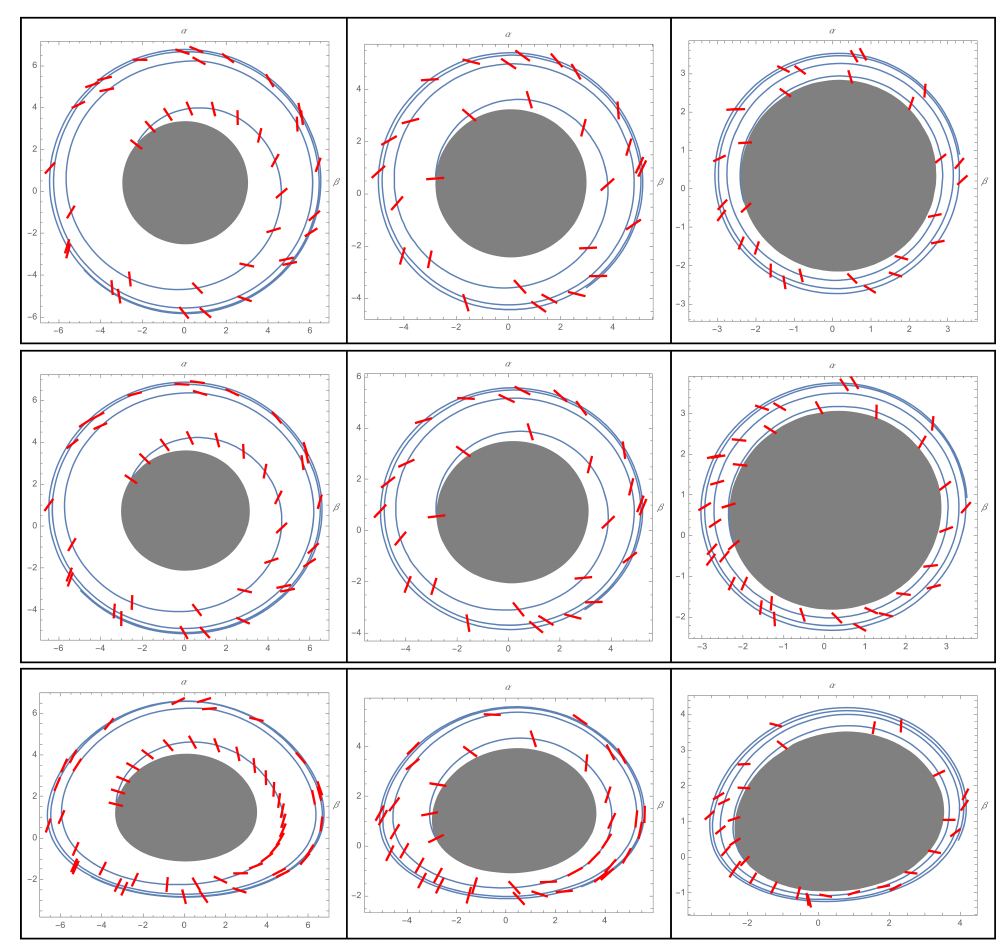}
	\centering
	\caption{The polarization structures of hotspots along a prograde plunging orbit under a vertical magnetic field. From left to right, the black hole spin is set to $a = 0.1M, 0.5M,$ and $0.9M$. From top to bottom, the inclination angle is set to $17 ^\circ, 30 ^\circ,$ and $60 ^\circ$.}
	\label{angles}
\end{figure}

In Figure \ref{angles}, we illustrate the polarization vectors along the projected trajectory of a plunging hotspot on the screen, assuming a vertical magnetic field. From left to right columns, the black hole spin is set to be $0.1M, 0.5M,$ and $0.9M$. From top to bottom rows, the inclination angle is set to be $17 ^\circ, 30 ^\circ,$ and $60 ^\circ$. As the inclination angle increases, the projection of the ISCO on the screen gradually transforms from a circle to an elliptical shape, leading to the corresponding changes in the appearance of the hotspot's image. It is worth noting that for small inclination angles, such as $\theta=17^\circ$ and $30^\circ$, the results exhibit only minor deviations from the one in the on-axis case. Given that current observations of M87 and Sgr A* suggest a preference for small inclination angles, our previous discussion for the on-axis case remains qualitatively valid.

\begin{figure}[th!]
	\centering
	\includegraphics[width=6.2in]{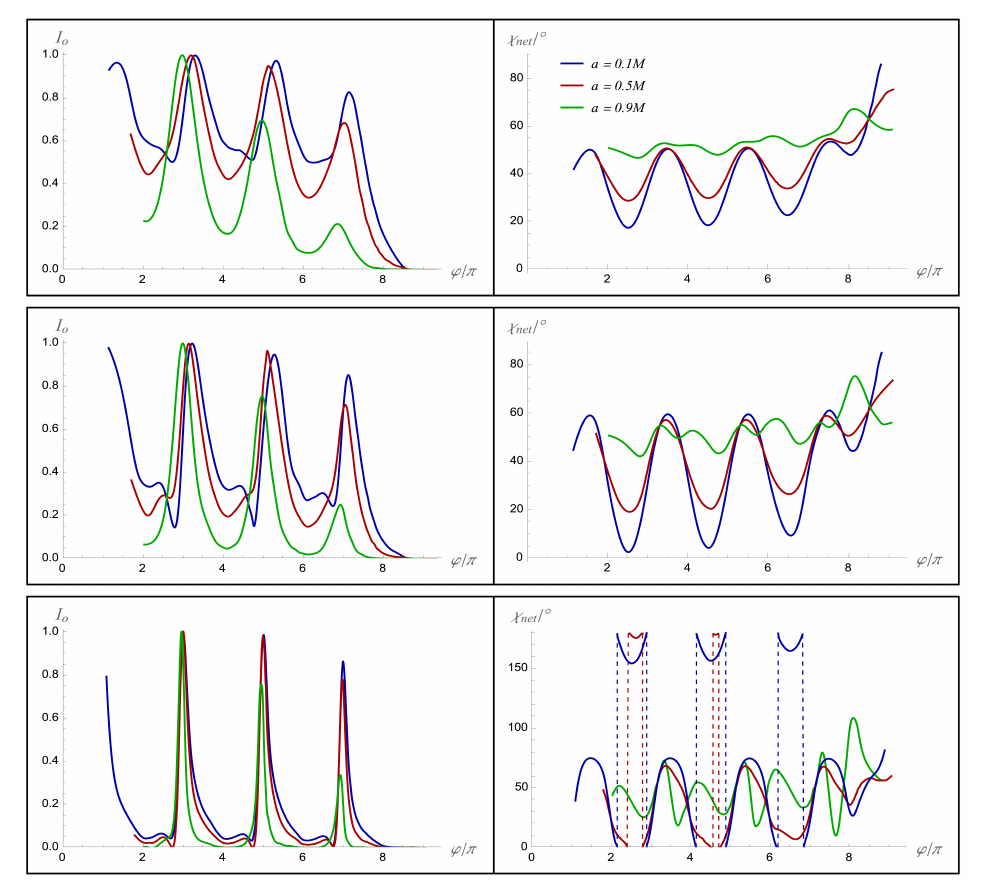}
	\centering
	\caption{The observed total intensity and net EVPA along a prograde plunging orbit under a vertical magnetic field. From top to bottom, the inclination angle is set to $17 ^\circ, 30 ^\circ,$ and $60 ^\circ$. The maximum value of the total intensity has been normalized to 1. The maximum value of $\varphi$ corresponds to $r_s=r_{+}+0.05M$ and the minimum value of $\varphi$ corresponds to $r_s=r_{ms}-0.1M$.}
	\label{angles_IE}
\end{figure}

In Figure \ref{angles_IE}, we illustrate the variations in the observed total intensity and net EVPA along the projected trajectory of the hotspot. To describe the projected trajectory, we introduce polar coordinates defined as $\rho = \sqrt{\a^2+\b^2}, \varphi = \arctan{(\b/\a)}$. For nonzero inclination angles, the angular radius $\rho$ along the projected trajectory on the observer’s screen is no longer a monotonic parameter, in contrast to the on-axis case. Consequently, we plot the observed quantities as the functions of the polar angle $\varphi$ on the screen. As the hotspot inspirals counterclockwise toward the black hole, $\varphi$ increases. A clear oscillatory behavior in both $I_o$ and $\chi_{net}$ with respect to $\varphi$ is evident, 
which can be attributed to the azimuthal angle $\phi$ completing multiple full $360^\circ$ rotations around the black hole.

In Figure \ref{angles_IE}, each peak in the oscillation of $I_o$ corresponds to a position near the left side of the $\alpha$-axis on the observer’s screen, resulting from the Doppler effect as the hotspot moves toward the observer. This effect becomes more pronounced with increasing inclination angle. Additionally, the variations in the observed quantities across different oscillation cycles reflect the hotspot’s gradual deviation from the ISCO. As the hotspot plunges into the black hole, the heights of the intensity peaks  progressively decrease, a trend that is more pronounced in high-spin cases. Regarding the polarization pattern, it generally retains a counterclockwise orientation. As the inclination angle increases, the oscillation amplitude of $\chi_{net}$ also grows. For an inclination of $60^{\circ}$, a portion of the polarization in the upper half of the screen exhibits a clockwise pattern. Moreover,  the variation in $\chi_{net}$ is smoothest when $a = 0.9M$, regardless of the inclination angle.

\subsection{Results for accretion flows within the plunging region}

In the intra-ISCO area of a thin accretion disk, the fluid inevitably adopts a plunging trajectory towards the black hole's horizon, forming a distinct plunging region. This region holds significant astrophysical implications \cite{Reynolds:2006uq, Zhu:2012vf, Mummery:2023tgh}. Notably, it plays a crucial role in explaining the observational spectrum of accretion disks in the X-ray band \cite{Machida:2002ub, Wilkins:2020pgu}. If the plunging region is present near a supermassive black hole like M87* or SgrA*, it can also yield unique signatures in the radio band, potentially detectable by the EHT. Within this plunging region, the fluid dynamics are primarily governed by the gravitational pull, causing the streamlines to closely follow timelike geodesics. The analytical framework outlined in Sec.\ref{theKerr} and Sec. \ref{imaginghs} can be directly applied to study the plunging region. For example, superimposing the trjectories of Eq.~\eqref{f012isco} with varying initial azimuthal angles can depict a stable, axisymmetric profile of an intra-ISCO accretion disk.

\begin{figure}[th!]
	\centering
	\includegraphics[width=6.2in]{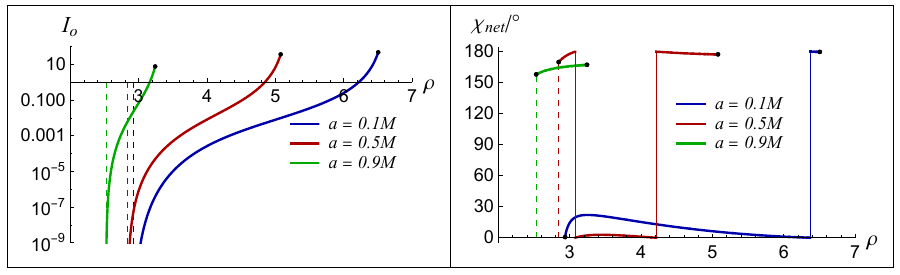}
	\centering
	\caption{The observed total intensity and the net EVPA for a plunging-region disk comprised of entirely zero-conductivity plasma. We set the radial velocity at the ISCO crossing as $\d{U}=10^{-3}$. For each case, the dashed line denotes the projection of the event horizon calculated using Eq.~\eqref{inverse1}.}
	\label{innerISCOdisk}
\end{figure}

In a typical accretion disk, the magnetic field is intricately linked with the fluid dynamics. In the case of a plasma with infinite conductivity, the accompanying magnetic field is governed by the ideal-MHD condition \cite{Hou:2023bep}, resulting in  
\bea
B^t = -\f{\Psi}{\sqrt{-g}} \bigg( U_r + \f{U^{\phi}}{U^r}L \bigg) \, , \quad B^r = -\f{\Psi}{\sqrt{-g}} E \, , \quad B^\phi = -\f{\Psi}{\sqrt{-g}} \f{U^{\phi}}{U^r} E \, .
\label{B}
\eea
Here $g$ is the determinant of the metric. Note that $B^r, B^\p$ diverge at the ISCO where $U^r = 0$. This divergence should be addressed by introducing a non-zero radial velocity when crossing the ISCO, as suggested by \cite{Mummery:2023tgh}, given by $U^{r}=U^{r}_{I} - \d U$, where $\d U \ll 1$ is a small correction based on Eq.~\eqref{f012isco}. In the zero-conductivity fluid, the magnetic field lines align with the streamlines, leading to a negligible vertical component near the equatorial plane. However, real accretion flows may exhibit some conductivity near the equatorial plane, resulting in a non-zero vertical component there. For simplicity, the Papapetrou-Wald potential given by Eq.~\eqref{Wald} can be utilized to represent a typical vertical magnetic field within the plunging region. Consequently, the results depicted in the top row of Fig.~\ref{intEVPA} can be interpreted as the polarization pattern of the plunging region with a vertical magnetic field. The suppression of central intensity outlines the horizon on the screen, commonly referred to as the inner shadow of the accretion disk \cite{dokuchaev2019event, Chael:2021rjo}. 
 
Under the accompanying magnetic field of Eq.~\eqref{B}, the image of the disk seen by an on-axis observer is illustrated in Fig. \ref{innerISCOdisk}, where we have chosen $\d{U}=10^{-3}$ as a conservative estimation. The behavior of $\mathcal{I}_o$ as the radius varies reveals a significant reduction in the observed intensity not so close to the horizon. For instance, $\mathcal{I}_o$ at $r = (r_++r_{I})/2$ diminishes to nearly $1/10000$ of the value at $r = r_{I}$. Thus, with such a magnetic field configuration, the plunging region appears much darker, resulting in a larger central dark region on the screen compared to the inner shadow produced by the vertical field. Regarding the polarization pattern, as the radius decreases from the ISCO, $\chi_{net}$ remains less than $\pi/2$ across all scenarios under the vertical magnetic field, as depicted in the top-right panel of Fig. \ref{intEVPA}, indicating a counterclockwise pattern. Conversely, with the accompanying magnetic field, $\chi_{net}$ is close to $0$ or $\pi$, indicating a highly radial polarization pattern, as shown in the right panel of. Fig. \ref{innerISCOdisk}. Furthermore, for $a = 0.1M$ or $0.5M$, the net EVPA undergoes a transition from weakly clockwise to weakly counterclockwise patterns, manifesting as jumps from $180^{\circ}$ to $0^{\circ}$. For $a = 0.9M$ the net EVPA exhibits a weakly clockwise pattern.

\subsection{Higher-order images}

In the preceding discussion, we did not consider the higher-order images. Here, we briefly explain why we neglect these lensed images. The higher-order images arise from the photons that undergo multiple orbits around the black hole before reaching the observer. These images are confined within the so-called lensing band, whose inner and outer boundaries correspond to the lensed images of circles at $r = r_+$ and $r \rightarrow \infty$ on the equatorial plane respectively\cite{Gralla:2019drh,Cardenas-Avendano:2023obg}. For small inclination angles, the lensing band is located very close to the critical curve, which is characterized by the photon-sphere radius \cite{bardeen1973black}:
\bea\label{tdr0}
&&\tilde{r}_0 = M + 2\sqrt{M^2-\f{a^2}{3}} \cos{\left[ \f{1}{3}\arccos{  \f{M\left(M^2-a^2\right)}{\left( M^2- \f{a^2}{3} \right)^{3/2}}  } \right]} \,, \nn \\
&&  \tilde{\rho}_0 = \sqrt{\f{\tilde{r}_0^3}{a^2}\left[ \f{4M(\tilde{r}_0-r_+)(\tilde{r}_0-r_-)}{(\tilde{r}_0-M)^2}-\tilde{r}_0 \right] + a^2} \,.
\eea
As the spin decreases from $a = M$ to $0$, $\tilde{r}_0$ increases from $2.414M$ to $3M$. The radius of the critical curve, given by $\rho = \tilde{\rho}_0$, is determined by the critical impact parameters and varies from $4.828M$ to $3\sqrt{3}M$. 

\begin{figure}[th!]
	\centering
	\includegraphics[width=6.6in]{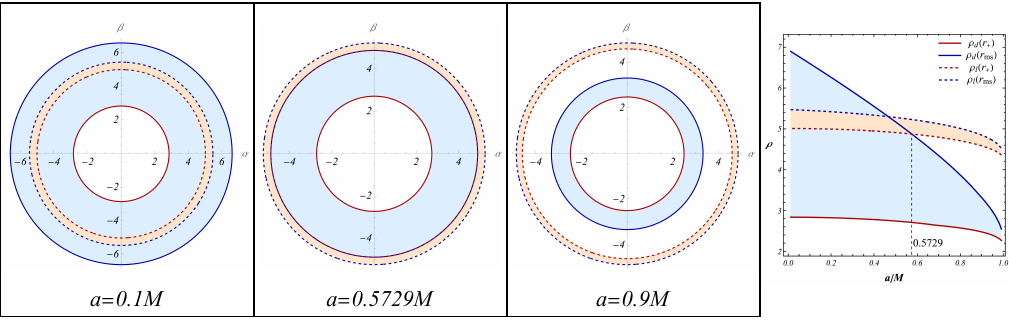}
	\centering
	\caption{The direct images and the higher-order images.  The blue curves represent the projections of the ISCO on the observer’s screen, while the red curves correspond to the projections of the event horizon. Solid and dashed curves indicate their direct and lensed images, respectively. Besides, we shaded the direct image region in light blue and the lensed image region in light yellow. From left to right, the black hole spin is set to $a = 0.1M, 0.5729M, 0.9M$. The rightmost panel illustrates the variations of $\rho_{d}(r_+),\rho_{d}(r_{ms}),\rho_{l}(r_+),\rho_{l}(r_{ms})$ as the functions of spin. The lensed images are computed using Eq. (78) in \cite{Gralla:2019drh}. When $a>0.5729M$, we have $\rho_{d}(r_+)<\rho_{d}(r_{ms})<\rho_{l}(r_+)<\rho_{l}(r_{ms})$, which means the lensed image is separated from the direct image.}
	\label{highorder}
\end{figure}

For the direct image, the angular radius follows the approximate relation $\rho \approx r/M + 1$ (see Eq.~\eqref{inverse2}) for small inclination angles, indicating that the direct image experiences minimal distortion. As the plunging orbit extends from the ISCO to the event horizon, the inner and outer boundaries of the time-integrated direct image are given by $\rho_{d}(r_+)  \approx r_+/M + 1$, $\rho_{d}(r_{ms})  \approx r_{ms}/M + 1$, where and hereafter we use $\rho_{d}(r)$, $\rho_{l}(r)$ to denote the angular radii of the direct and lensed images, respectively, on the observer’s screen. As the spin decreases from $a = M$ to $0$, $\rho_{d}(r_+)$ ranges from $2M$ to $3M$, while $\rho_{d}(r_{ms})$ varies from $2M$ to $7M$. Consequently, as the spin decreases, $\rho_{d}(r_{ms})$ decreases faster  than $\tilde{\rho}_0$. If the spin is sufficiently high such that $\rho_{d}(r_{ms}) < \tilde{\rho}_0 \approx \rho_{l}(r)$, then the time-integrated direct and higher-order images remain spatially separated on the screen. A precise calculation shows that the critical case occurs at $a = 0.5729M$, where we have $\rho_{d}(r_{ms}) = \rho_{l}(r_+)$. Therefore, as long as $a>0.5729M$, the time-integrated direct and higher-order images can be analyzed independently, as illustrated in Figure \ref{highorder}. 

Furthermore, we note that the  emission from the hotspot is inherently a dynamical process. In astrophysics, the typical timescale for plunging objects in Kerr black holes is approximately $t_{\text{pl}} = 2M \sim 5M$ \cite{Mummery:2023tgh}. In contrast, the time delay between the photons forming adjacent-order images is given by $t_{\text{dl}} \approx 3\sqrt{3}\pi M$ \cite{Gralla:2019drh}. Since $t_{\text{dl}} > t_{\text{pl}}$, the direct image of the plunging hotspot will be observed before the lensed photons for higher-order images reach the observer. Consequently, the direct and higher-order images of the plunging hotspot remain separated in the time domain.

\section{Summary} \label{sum}

We studied the polarized images of luminous hotspots on non-circluar orbits, in close proximity to the horizon of a Kerr black hole. We focused on the synchrotron emission from the hotspots moving along the plunging orbits and the homoclinic orbits \cite{Mummery:2022ana, Levin:2008yp}, while considering two distinct magnetic field configurations: the vertical and radial fields \cite{Wald:1974np, Hou:2023bep}. For an on-axis observer, we introduced a new approximate function to facilitate the computation of the light bending, thereby simplifying the calculation of the imaging process. To elucidate the polarization patterns, we conducted the calculations involving the Stokes parameters and derived the total intensity and the polarization vector along a hotspot's trajectory. The main results of our study are listed below:
\begin{itemize}
\item During the inspiral of a hotspot from the prograde ISCO, when subject to a vertical magnetic field, the total intensity of the hotspot decreases monotonically as it approaches the horizon, while the net EVPA continually increases but remains $\chi_{net} < \pi/2$. In contrast, with a radial magnetic field, the total intensity does not follow monotonic patterns, and the net EVPA always satisfies $\chi_{net} > \pi/2$. The Stokes parameters demonstrate a right-handed spiral shape on the Q-U plane, with the phase determined by the hotspot's motion in the $\p$ direction. 

\item When a hotspot plunges from the retrograde ISCO, the projected trajectory on the observer's screen can be divided into two parts: one segment inspiraling retrogradely, and the other inspiraling progradely. The retrograde segment appears significantly brighter, with a net EVPA greater than $\pi/2$, contributing to nearly the entire Q-U curve. In contrast, the prograde segment is notably dim, to the extent that it is scarcely discernible in the Stokes parameters, as demonstrated in Fig. \ref{retro}. The net EVPA of the prograde segment is less than $\pi/2$.   

\item When a hotspot travels along a homoclinic orbit, the resulting Q-U curve showcases a novel multi-loop pattern, as depicted in Fig. \ref{homo}. It is noteworthy that while a Q-U one loop is a common characteristic of elliptical-like orbits, it is exclusively the homoclinic orbit that can produce a multi-loop structure on the Q-U plane. This phenomenon arises from the hotspot continuously encircling an unstable circular orbit.

\item For a nonzero inclination angle, the time-integrated image takes on an elliptical shape, with the total intensity significantly influenced by the Doppler effect. The polarization pattern is generally counterclockwise; however, for large inclination angles ($\gtrsim 60^{\circ}$), a portion of the net EVPA in the upper half of the screen exhibits a clockwise pattern. Additionally, a higher spin more effectively smooths out the variations in the net EVPA along the trajectory.
\end{itemize}

Moreover, by considering the plunging region of an accretion flow approximated by plunging geodesics, we extended our study to the polarized images of the intra-ISCO region of a thin disk. Our study provided, for the first time, the details of the polarizations of a plunging region disk under different magnetic field configurations. Under the vertical magnetic field or the accompanying magnetic field given by Eq.~\eqref{B}, the images of the disk are totally different, as illustrated in Fig. \ref{intEVPA} and Fig. \ref{innerISCOdisk}. Upon comparing the total intensities between the two scenarios, we observed that the plunging region with the accompanying magnetic field appears notably darker than when under the vertical magnetic field.  Furthermore, the net EVPA under the vertical magnetic field indicates a counterclockwise pattern, whereas the net EVPA under the accompanying magnetic field exhibits a highly radial pattern.

We end this work by outlining several limitations and future perspectives of present study. In order to carry an analytical study, we only considered the hotspots traveling along the plunging and the homoclinic geodesics with integrable trajectory functions. It is essential to consider the hotspots with other types of motions, such as the bound orbits and the deflecting orbits \cite{Cieslik:2023qdc}, or the hotspots not confined to the equatorial plane. Additionally, we have neglected the effects of higher-order images. In the high-spin cases, the time-integrated higher-order images are well separated from the direct one and do not significantly affect the results. However, in the low-spin cases, as there are overlap between the time-integrated direct and higher-order images, one must take the higher-order images seriously. Moreover, the higher-order images may contribute to the polarized light curve as well, and they should be treated with care. We would like to leave these issues to future studies.

\section*{Acknowledgments}

The work is partly supported by NSFC Grant No. 12275004. 

\bibliographystyle{utphys}
\bibliography{plrzhspt}
		
\end{document}